\documentclass[11pt]{article}

\usepackage[english]{babel}
\usepackage{amssymb}
\usepackage{amsfonts}
\usepackage{cite}
\usepackage{bbold}

\usepackage{amssymb,amsmath,slashed}
\usepackage{epsf}

\renewcommand{\baselinestretch}{1.3}
\makeatletter
\def\@begintheorem#1#2{\trivlist%
 \item[\hskip \labelsep{\bfseries #2\ #1}]\itshape}
\newtheorem{teo}{Theorem}[section]

\newtheorem{_rem}[teo]{Remark}
\newtheorem{_rems}[teo]{Remarks}
\newtheorem{_eje}[teo]{Example}

\makeatother

\newenvironment{Proof}{{\em Proof:}}{\hfill $\square$
\vspace{3mm}}

\DeclareMathAlphabet{\Ma}{U}{msa}{m}{n}
\DeclareMathAlphabet{\Mb}{U}{msb}{m}{n}
\DeclareMathAlphabet{\Meuf}{U}{euf}{m}{n}

\def\got#1{\Meuf{#1}}

\DeclareSymbolFont{ASMa}{U}{msa}{m}{n}
\DeclareSymbolFont{ASMb}{U}{msb}{m}{n}
\DeclareMathSymbol{\hrist}{\mathord}{ASMa}{"16}
\DeclareMathSymbol{\varkappa}{\mathalpha}{ASMb}{"7B}
\DeclareMathSymbol{\CrPr}{\mathord}{ASMb}{"6F}

\def\restriction{\upharpoonright}

  \def\al #1.{{\mathcal{#1}}}
  \def\ot #1.{{\got{#1}}}

  \def\ccr #1,#2.{\overline{\Delta(#1,\,#2)}}

  \def\b #1.{{\bf #1}}
  \def\cross#1.{\mathrel{\mathop{\times}\limits_{#1}}}
  
  \def\C{\Mb{C}}

  \def\R{\Mb{R}}
  
 \def\un{{\mbox{\boldmath$1$}}}

\def\f #1,#2.{\mathsurround=0pt \hbox{${#1\over #2}$}\mathsurround=5pt}

  \def\cross #1.{\mathrel{\raise 3pt\hbox{$\mathop\times\limits_{#1}$}}}
  \def\ol #1.{\overline{#1}}
\def\b #1.{{\bf #1}}

\def\ker{{\rm Ker}\,}

\def\ran{{\rm Ran}\,}

\def\Her #1.{{\rm Her}(#1)}

\def\s #1.{_{\smash{\lower2pt\hbox{\mathsurround=0pt $\scriptstyle #1$}}\mathsurround=5pt}}
\def\set #1,#2.{\left\{\,#1\;\bigm|\;#2\,\right\}}
\def\maprightu #1;{\smash{\mathop{\longrightarrow}\limits^{#1}}}
\def\maprightd #1;{\smash{\mathop{\longrightarrow}\limits_{#1}}}
\def\maprightt #1,#2.{\mathrel{\smash{\mathop{\longrightarrow}\limits_{#1}^{#2}}}}

\def\XP#1!{\renewcommand{\baselinestretch}{.7}\marginpar{$\leftarrow${\footnotesize #1}\hfil}
 \renewcommand{\baselinestretch}{1}}
\def\XB{\marginpar{
{\footnotesize Change~starts-}\lower 11pt\hbox{\mathsurround=0pt$
\!\!\displaystyle{
\Bigg\downarrow}$\mathsurround=3pt}}}
\def\XE{\marginpar{{\footnotesize Change~ends-}\raise 10pt\hbox{\mathsurround=0pt$
\!\!\displaystyle{
\Bigg\uparrow}$\mathsurround=3pt}}}
\def\f #1,#2.{\mathsurround=0pt \hbox{${#1\over #2}$}\mathsurround=5pt}

\def\margin #1.{\marginpar{#1}}

\def\lcs{Lie C*--system}
\def\lcss{Lie C*--systems}

\newtheorem{Definition}{Definition}[section]
\newtheorem{Theorem}[Definition]{Theorem}
\newtheorem{Proposition}[Definition]{Proposition}
\newtheorem{Lemma}[Definition]{Lemma}

\newcommand{\ie}{{\it i.e.\ }}
\newcommand{\eg}{{\it e.g.\ }}

\newcommand{\RR}{\mathbb R}
\newcommand{\CC}{\mathbb C}
\newcommand{\NN}{\mathbb N}

\newcommand{\cA}{{\cal A}}

\newcommand{\cD}{{\cal D}}
\newcommand{\cAo}{{\cal A}_0}
\newcommand{\cH}{{\cal H}}
\newcommand{\cK}{{\cal K}}
\newcommand{\cR}{{\cal R}}
\newcommand{\cBH}{{\cal B(H)}}
\newcommand{\cDf}{{\cal D}_f}

\newcommand{\delf}{\delta_f}
\newcommand{\delg}{\delta_g}

\newcommand{\Rlf}{R(\lambda,f)}

\newcommand{\Rmg}{R(\mu,g)}

\newcommand{\Rplf}{R{\, ^\prime}(\lambda,f)}

\newcommand{\Gf}{G_f}
\newcommand{\Gg}{G_g}

\newcommand{\Gpf}{G^{\, \prime}_f}

\title{\bf Lie Algebras of Derivations and Resolvent Algebras}
\author{\large {Detlev Buchholz\,$^a$\thanks{Supported by the German
Research Foundation (Deutsche Forschungsgemeinschaft (DFG)) through the
Institutional Strategy of the University of G\"ottingen}
 \ and \
Hendrik Grundling\,$^b$}\\[4mm]
${}^a$ Institut f\"ur Theoretische Physik and
Courant Centre \\
``Higher Order Structures in Mathematics'',
Universit\"at G\"ottingen, \\ 37077 G\"ottingen, Germany  \\[2mm]
${}^b$   Department of Mathematics,  University of New South Wales \\
Sydney, NSW 2052, Australia  \\[10mm]
Dedicated to the memory of Hans--J\"urgen Borchers
}

\date{}

\begin{document}
\maketitle
\begin{abstract}
\noindent
This paper analyzes the action  $\delta$ of a Lie algebra $X$ by 
derivations on a C*--algebra $\cA$. This action satisfies                      
an ``almost inner'' property which 
ensures affiliation of the generators of the derivations
$\delta$ with $\cA$, and is expressed in terms of corresponding
pseudo--resolvents.
In particular, for an abelian Lie algebra $X$ acting on a primitive 
C*--algebra $\cA$, it is shown that there is a central extension 
of $X$ which determines algebraic relations of the underlying
pseudo--resolvents.
If the Lie action $\delta$ is ergodic, \ie the 
only elements of $\cA$ on which all the derivations 
in $\delta_X$  vanish are multiples of the identity, then
this extension is given by a (non--degenerate) 
symplectic form $\sigma$ on $X$. Moreover, the algebra
generated by the pseudo--resolvents coincides with
the resolvent algebra based on the symplectic space
$(X,\sigma)$. Thus the  resolvent algebra of the 
canonical commutation relations,  which was recently
introduced in physically motivated analyses of quantum
systems, appears also naturally 
in the representation theory of
Lie algebras of derivations acting on C*--algebras. 
\end{abstract}

\section{Introduction and Framework}
\setcounter{equation}{0}
In quantum physics, symmetry transformations are often given in terms of 
their infinitesimal
action on the algebra of observables, \ie 
as a Lie action of derivations on the observables.  
It is not always clear that there is a faithful representation of 
the observables in which the Lie algebra is represented by 
(selfadjoint) operators implementing the given 
Lie action by commutators.
In fact, such covariant representations
may not exist and one needs to take a cocycle representation 
of the Lie algebra to obtain the implementing property, 
or equivalently replace 
the Lie algebra by a central extension of it.
Two prominent examples are first,  the abelian group of position and velocity
transformations in quantum mechanics and second,  the
conformal transformations in two--dimensional quantum
field theory. For these, the central 
extensions are the Heisenberg algebra and the Virasoro
algebra, respectively. 
Which central extensions appear seems to be 
fixed by the structure of the underlying algebra~$\cA$.
This fact has been observed in the context of quantum
anomalies in many examples; but there does not yet
exist a systematic investigation of it.
It is the aim of the present article to
begin such a study in a setting based on the
following assumptions.

\vspace*{2mm} \noindent
\textbf{(I)} \
Let $X$ be a (finite or infinite dimensional)
real Lie algebra with Lie--bracket
$[ \cdot, \cdot ]$ and let $\cal A$ be a unital C*-algebra which
is primitive (\ie it has a faithful irreducible representation).
Let ${\cal A}_0 \subset \cal A$ be a norm dense unital *-subalgebra,
and let $\delta:X\to{\rm Der} {\cal A}_0$ be an injective Lie homomorphism
into the Lie algebra ${\rm Der} {\cal A}_0$ of *-derivations
of ${\cal A}_0$, \ie $\delta$ is real linear and
\begin{equation} \label{1.1}
\delf \circ \delg - \delg \circ \delf
= \delta_{[f,g]} \, ,\;\; f,g\in X.
\end{equation}
Such a pair $(X, {\cal A})$ will be called a \textit{\lcs}.
The action $\delta$ is said to be {\it ergodic} if
$\delta_f(A_0) = 0$ for all $f \in X$ implies that
$A_0$ is a multiple of the identity.

\vspace*{2mm}
This framework  covers quantum physics, where algebras of
observables are generally constructed in some distinguished irreducible
representations, \eg the Fock representation. 
On the other hand,
it excludes classical physics, where the observable
algebras are abelian and symmetries act in a non--trivial manner.
The intermediate cases, where the underlying algebras of observables
have a center on which $\delta$ acts trivially,
such as in the presence of superselected charges,
can usually be reduced to the present setting by
proceeding to suitable quotient algebras.

Of particular interest for physics are \lcss \
where the action $\delta$ is induced by selfadjoint
generators which can be interpreted as
observables.  The simplest case
is  if the derivations $\delta_f$ are inner,
\ie if for  each $f \in X$ there are
operators $G_f = G_f{}^* \in {\cal A}$ such that
$\delta_f(A_0) = i \, [G_f, A_0 ]$ for all $A_0 \in {\cal A}_0$.
However, generically the generators of symmetries are
unbounded operators and hence are not elements of
the underlying C*--algebra. In order to see how
to deal with these cases we rewrite the
preceding equation in terms of the
resolvents of the generators $G_f$:
\[
(i \lambda \un + G_f)^{-1} \delta_f(A_0) (i \lambda \un +
G_f)^{-1}  = i [A_0 ,(i \lambda \un + G_f)^{-1} ]
\]
for
$\lambda \in \RR \backslash \{ 0 \}$. This equation
can be generalized so as to cover the case of unbounded generators
which are affiliated with $\cA$ by making use of the
notion of pseudo--resolvent \cite{Yo}; 
it replaces the familiar concept of the 
resolvent of a selfadjoint operator in the abstract C*--setting.
\\[2mm] 
\textbf{Definition:} \ Let $\al A.$ be a C*-algebra.
A {\it pseudo--resolvent} is a function
$R : \RR \backslash \{ 0 \}  \rightarrow  {\cal A}$
such that
$$
R(\lambda) - R(\mu) = i(\mu -\lambda)
R(\lambda)R(\mu) \, , \quad
R(\lambda)^*=R(-\lambda)
\quad \hbox{for}\quad \lambda, \mu \in\RR \backslash \{ 0 \}.
$$
Any pseudo--resolvent can be analytically
continued to the domain $\C\backslash{i\R}$.
By some abuse of terminology, we will use the term
pseudo--resolvent also for its
values $R(z)$,  $z \in \C\backslash{i\R}$.

\vspace*{2mm} 
\noindent 
With this concept we
can express the assumption that there
exist (possibly unbounded) generators of the action $\delta$
which are affiliated with the algebra~${\cal A}$.

\pagebreak 
\noindent \textbf{(II)} \
Let $(X,{\cal A})$ be a \lcs. The underlying
action $\delta$ is said to be \textit{almost inner} if,
for each $f \in X$, there is a pseudo--resolvent
$\Rlf \in {\cal A}$,  $\lambda \in \RR \backslash \{ 0 \}$ such that
\begin{equation} \label{1.2}
\Rlf \, \delf(A_0) \, \Rlf = i \, [A_0 , \Rlf] \, , \quad A_0 \in \cAo \, .
\end{equation}

\vspace*{1mm} 
\noindent
\textbf{Remarks:} \ It can be seen that 
relation (\ref{1.2}) holds for all values
of $\lambda$ if it holds for one. As a matter of fact, by
analytic continuation in $\lambda$, it holds on the
entire domain $\C\backslash{i\R}$. Moreover, the relation
implies that $R(\lambda,0)$ is contained in the center
of ${\cal A}$ and hence must be a multiple of
$\un$ since ${\cal A}$ is primitive. Assuming that this
multiple is different from $0$, it follows
from the defining relations of pseudo--resolvents that one
can consistently put $R(\lambda,0)=-(i/\lambda) \un$,
$\lambda \in \RR \backslash \{ 0 \}$.

\vspace*{1mm}
We will show below that condition (II) implies that 
$\delta$ is induced by selfadjoint generators in every faithful
irreducible representation of ${\cal A}$. This observation is the
key to analyzing the algebraic properties of the
pseudo--resolvents inherited from the Lie structure of the
derivations. For this analysis we need the
following technical assumption which holds
only for a restricted class of Lie algebras $X$, such as
compact or abelian ones.

\vspace*{1mm} \noindent
\textbf{(III)} \ Let $(X,{\cal A})$ be a \lcs{}
 satisfying (II). For each $f \in X$, there is a pseudo--resolvent 
satisfying relation~(\ref{1.2}) which 
is in the domain ${\cal A}_0$, \ie
$R(z,f) \in {\cal A}_0$ for  all $z \in \CC \backslash i \RR$.

\vspace*{1mm}
Having stated the general framework, we now restrict
the subsequent analysis to
the simple but physically important case of abelian Lie algebras $X$.
In this case the
action $\delta$ is flat, \ie the
right hand side of equation (\ref{1.1}) vanishes, and this 
explains the terminology used in the following definition.\\[1mm]
\textbf{Definition:} \ Let $X$ be a real abelian Lie algebra.
The pair $(X, {\cal A})$ is said to be a \textit{flat \lcs}
if it satisfies the conditions (I), (II) and (III).

\vspace*{1mm}
Given any flat \lcs{}
$(X, {\cal A})$, we will determine the structure of the algebra
$\cR \subset \cA$ generated by the associated
pseudo--resolvents. It contains information about
the commutation relations of the generators which
implement the action $\delta$ and hence about the
possible appearance of central extensions of $X$. 
We will see that for any such system there is a unique
skew symmetric bilinear form $\sigma : X \times X \rightarrow \RR$
fixing  an extension.
If $\delta$ acts ergodically
on $\cal A$, the form~$\sigma$ is non--degenerate
and $(X, \sigma)$ is a symplectic space.
The algebra $\cR$ then coincides with the resolvent algebra
$\cR(X, \sigma)$, defined in \cite{BuGr1}.
Moreover, if $X$ is finite--dimensional,
then its dimension must be even and the algebra $\cal R$
is the unique Heisenberg algebra of canonical
commutation relations in resolvent form.

The article is organized as follows.
The basic framework and assumptions are specified in this introduction.
In Sect.~2 we work out the algebraic consequences, and establish existence of
a skew symmetric bilinear form $\sigma$ on $X \times X$
entering into the commutation relations of the pseudo--resolvents.
In Sect.~3 we show by standard cohomological arguments
that the pseudo--resolvents can be chosen in such a way
that they also encode linearity of the action $\delta$ on $X$.
We obtain therefore all the defining relations of the resolvent
algebra on $(X, \sigma)$
and hence this algebra is a subalgebra of $\cA$. We also show
uniqueness of this subalgebra relative to the initial action.

\section{Algebraic structure}
\setcounter{equation}{0}

Henceforth, we will assume that $(X, {\cal A})$
is a flat \lcs.
As $\cA$ is a primitive unital C*--algebra,
it has by definition a faithful irreducible representation.
Thus we may assume without loss of generality that we
have concretely $\cA\subseteq \cBH$ for some
Hilbert space $\cH$ and $\cA^- = \cBH$, where the bar denotes
weak closure.
\begin{Lemma} \label{l2.1}
Let $f \in X$ and $\lambda \in  \RR \backslash \{0\}$.
\begin{itemize}
\item[(i)] There is a selfadjoint operator (generator)
 $\Gf$ with domain $\cDf = \Rlf \, \cH$ such that
$\Rlf = {(i\lambda \un + G_f)^{-1}}$.
\item[(ii)] For each $A_0 \in \cAo$ there are 
$B_0, C_0 \in \cAo$ such that
$$ A_0 \, \Rlf = \Rlf \, B_0 \ \ \mbox{and} \ \ \Rlf \, A_0 = C_0 \, \Rlf \, .$$
\end{itemize}
\end{Lemma}
\begin{Proof}
(i)
From relation (\ref{1.2}) we see that $\ker\Rlf \subset \cH$, 
the kernel of $\Rlf$,
is stable under the action of $\cAo$ and,
by continuity, also under the action of $\cA$. As $\cA$
is irreducible, the kernel can only consist of $\{0\}$.
However $\ker\Rlf=\{0\}$ iff $\Rlf$ is a resolvent
 $\Rlf=(i\lambda+G_f)^{-1}$ by \cite[Corollary 1]{Ka59},
and $G_f$ has domain $\cDf = \Rlf \, \cH$.
Now $G_f$ is symmetric by
$$
G_f^*=\big(R(1,f)^{-1} - i\un
\big)^*\supseteq i\un+\big(R(1,f)^{-1}  \big)^*= i\un+R(-1,f)^{-1}
=G_f \, .
$$
That it is also selfadjoint follows from the equality of ranges
$\ran(\pm i \un +G_f)^{-1}=\ran R(\pm 1,f)$, the latter being dense
by \cite[Theorem on p.\ 467]{Ka59} since $\ker\Rlf=\{0\}$. \\[1mm]
(ii) It folows from relation (\ref{1.2}) that
$$
A_0 \, \Rlf = [A_0, \Rlf] + \Rlf A_0
= \Rlf \, \big(\!-i \delf(A_0) \, \Rlf + A_0 \big) \, .
$$
But $B_0 \doteq  \big(\!-i \delf(A_0) \, \Rlf + A_0 \big) \in \cAo$
since, by assumption, $\cAo$ is stable under the action of the
derivations and the resolvents are elements of this algebra
by condition (III).
This proves the first part of the statement; the second part
follows by a similar argument.
\end{Proof}

Note that $\cDf = \Rlf \, \cH$ does not depend on~$\lambda$.
It is also noteworthy that the first part of the lemma holds
for arbitrary \lcss{} satisfying (II), only in the second part
did we use assumption (III).
From this lemma we  obtain:

\begin{Lemma} \label{l2.2}
Let $f \in X$. Then $\cAo \, \cDf \subset \cDf$ and
\begin{equation} \label{2.1}
[\Gf , A_0] \, \Psi = -i \delf(A_0) \, \Psi \, , \quad A_0 \in \cAo, \, \Psi
\in \cDf \, .
\end{equation}
\end{Lemma}
\begin{Proof} It follows from part (ii) of the preceding lemma that
$\cAo \, \Rlf \, \cH \subset  \Rlf \, \cH$, proving the
stability of the domain $\cDf$ under the action of $\cAo$.
Now let $A_0 \in \cAo$ and $\Psi\in\cDf$, \ie
 $\Psi = \Rlf \Phi$ for some $\Phi \in \cH$. Then,
using computations in the preceding lemma,
\begin{equation*}
\begin{split}
& \Gf A_0 \Psi + i \lambda A_0 \Psi \\
& = (i\lambda 1 + \Gf) A_0 \Rlf \Phi \\
& = (i\lambda 1 + \Gf) \Rlf \, \big(\!-i \delf(A_0) \, \Rlf + A_0
\big) \Phi \\
& = \big(\!-i \delf(A_0) \, \Rlf + A_0 \big) \Phi \\
& = \big(\!-i \delf(A_0) + A_0 (i\lambda 1 + \Gf)  \big) \Rlf \Phi \\
& = \big(\!-i \delf(A_0) + A_0 \Gf \big) \Psi
+ i \lambda A_0 \Psi \, ,
\end{split}
\end{equation*}
proving relation (\ref{2.1}).
\end{Proof}

To analyze the commutation relations between the generators
$\Gf, G_g$, one needs more detail about their domains. The following
result provides the relevant information.

\begin{Lemma}  \label{l2.3}
\ Let $f,g \in X$ and let $\lambda, \mu \in \RR \backslash \{0\}$.
Then
$ \cD_{f,g,f} \doteq \Rlf \Rmg \Rlf \, \cH$
is dense in $\cH$. Moreover,
$\cAo \, \cD_{f,g,f} \subset \cD_{f,g,f}$ and
$\cD_{f,g,f}$ is contained in the domains of $\Gf$, $\Gg$ as well as
of their products in either order.
\end{Lemma}
\begin{Proof} Since the resolvents are bounded and have dense
range it follows that $\cD_{f,g,f}$ is dense.
Next, by threefold application of Lemma \ref{l2.1},
there exists for any $A_0 \in \cAo$ some  $B_0 \in \cAo$
such that $A_0  \Rlf \Rmg \Rlf = \Rlf \Rmg \Rlf B_0$,
proving the stability of $\cD_{f,g,f}$ under the action of
$\cAo$. Finally, it follows from its very definition that
$\cD_{f,g,f}$ lies in the domains of $\Gf$ and $\Gg \Gf$;
for the proof of the remaining assertion one makes use again
of Lemma \ref{l2.1} which implies, bearing in mind that
$ A_0 \doteq \Rlf  \in \cAo$ according to condition (III),
$$
\Rlf \Rmg \Rlf = \Rmg B_0 \Rlf = \Rmg \Rlf C_0
$$
for certain specific operators $B_0, C_0 \in \cAo$.
Hence $\cD_{f,g,f} \subset \Rmg \Rlf \, \cH$ also lies in the domains
of $\Gg$ and $\Gf \Gg$, completing the proof of the statement.
\end{Proof}

Below in Lemma~\ref{l3.1} we will have to establish a stronger
version of this lemma. Making use of the above
result one can now compute
the commutator of $[\Gf, \Gg]$ with the elements of $\cAo$.
\begin{Lemma}  \label{l2.4}
Let $\Phi \in \cD_{f,g,f}$ and let $A_0 \in \cAo$. Then
$$
[\Gf, \Gg] \, A_0 \, \Phi = A_0 \, [\Gf, \Gg] \, \Phi \, .
$$
\end{Lemma}
\begin{Proof}
The following computation relies on the preceding lemmata:
\begin{equation*}
\begin{split}
& \Gf \Gg A_0 \Phi = G_f \big(\! -i \delg(A_0) + A_0 \, \Gg \big) \Phi \\
& = \big(\! -\delf \circ \delg(A_0) -i \delg(A_0) \Gf -i \delf(A_0) \Gg
+ A_0 \, \Gf \Gg \big) \, \Phi \, .
\end{split}
\end{equation*}
Interchanging $f$ and $g$ one obtains an analogous equality.
By subtraction  one arrives at
$$
[\Gf, \Gg] \, A_0 \Phi =
- (\delf \circ \delg - \delg \circ \delf)(A_0) \, \Phi +
A_0 \, [\Gf, \Gg] \, \Phi \, .
$$
But $\delf \circ \delg - \delg \circ \delf = 0$, completing the proof.
\end{Proof}

It follows from this result and the assumption that
$\cA$ is irreducible that the commutator of $\Gf, \Gg$
is a c--number. The argument is based on a generalization of
Schur's lemma adapted to unbounded operators.
As it will be applied at various places,
we recall here the well--known proof.
\begin{Lemma} \label{l2.5}
Let $f,g \in X$. There  is a constant
$\sigma(f,g)\in\R$, antisymmetric in $f,g$,
such that
$$
 [\Gf, \Gg] \Phi  =
i \sigma(f,g) \,  \Phi  \quad\hbox{for all}\quad\Phi \in \cD_{f,g,f}\,.
$$
\end{Lemma}
\begin{Proof}
Let $K \doteq -i \, [\Gf, \Gg] $ on the domain $\cD_{f,g,f}$.
According to the preceding lemma we have
$\langle \Psi, A_0 K \Phi \rangle = \langle \Psi, K A_0 \Phi \rangle$ 
for all $A_0 \in \cAo $ and $\Phi,\,\Psi\in \cD_{f,g,f}$.
Taking also into account Lemma \ref{l2.3} and
the fact that the generators are selfadjoint, we can proceed to
$\langle \Psi, K A_0 \Phi \rangle = \langle K \Psi, A_0 \Phi \rangle$.
As $\cAo$ is dense in $\cA$ this implies that
$$
\langle K \Psi, A \Phi \rangle  = \langle \Psi, A K \Phi \rangle
\,,\qquad A \in \cA,\; \Phi, \Psi\in\cD_{f,g,f} \,.
$$
 Now as $\cA \subset \cBH$ is irreducible, hence algebraically
irreducible \cite{Sa},  for any given one--dimensional
projection  $E \in \cBH$ and finite dimensional subspace
$\cK \subset \cH$ there exists an operator $A_{E, \cK} \in \cA$ which acts like
$E$ on $\cK$, \ie $A_{E, \cK}\restriction\cK=E\restriction\cK$,
cf. \cite[Theorem~2.8.3(i)]{Dix}.
 So one can replace in the above equality
the operator $A$ by the projections onto the rays of $\Psi$ and
$\Phi$, respectively (assuming $\Psi, \Phi \not=0$), giving
$$
\langle \Psi, K  \Phi \rangle =
\langle \Psi, \Phi \rangle \langle \Psi, K  \Psi \rangle / \|  \Psi
\|^2  =
\langle \Psi, \Phi \rangle \langle \Phi, K  \Phi \rangle / \|  \Phi
\|^2 \, .
$$
As $K$ is a symmetric operator,
it follows from this equality that
$ \sigma(f,g) \doteq \langle \Phi,K \Phi \rangle / \|  \Phi \|^2 $
is real and does not depend on the choice of $\Phi \in \cD_{f,g,f}$.
Hence $\langle \Psi,  K \Phi \rangle =
\sigma(f,g) \langle \Psi, \Phi \rangle$ and,
since $\cD_{f,g,f}$ is dense, this implies that
$K\Phi=\sigma(f,g) \Phi$ for $\Phi\in\cD_{f,g,f}$, as claimed.
The antisymmetry of $ \sigma(f,g)$ in $f,g$
follows from its definition.
\end{Proof}

This result allows us to establish
corresponding algebraic properties of the resolvents.
\begin{Lemma} \label{l2.6}
Let $f, g \in X$. Then
$$
[\Rlf, \Rmg] =  i \sigma(f,g) \, \Rlf \Rmg^2 \Rlf \, , \quad
\lambda, \mu \in \RR \backslash \{0\} \, .
$$
Furthermore,
$$
 \delta_f(\Rmg) = \sigma(f,g) \, \Rmg^2 \, , \quad \mu  \in \RR \backslash \{0\} \, .
$$
\end{Lemma}
\begin{Proof} Let $\Phi \in \cD_{f,g,f}$. Then, by
condition (III) and
Lemma \ref{l2.3}, $\Rmg \Rlf \Phi \in \cD_{f,g,f}$ and one can compute
\begin{equation*}
\begin{split}
& i \sigma(f,g) \, \Rlf \Rmg^2 \Rlf \, \Phi \\
& = \Rlf \Rmg \, [\Gf, \Gg] \, \Rmg \Rlf \, \Phi \\
& =  \Rlf \Rmg \, [(i\lambda 1 + \Gf), (i \mu 1 + \Gg)] \, \Rmg \Rlf \, \Phi \\
& = [ \Rlf, \Rmg] \, \Phi \, .
\end{split}
\end{equation*}
Since $\cD_{f,g,f}$ is dense in $\cH$ the first part of the statement follows.
For the proof of the second part one makes use of
condition (III), the preceding result and Lemma \ref{l2.2}, giving
\begin{equation*}
\begin{split}
& -i \delf(\Rmg) \Rlf \, \Phi \\
& = (i \lambda 1 +  \Gf ) \Rmg \Rlf \, \Phi -  \Rmg \, \Phi \\
& = (i \lambda 1 +  \Gf ) \Rlf \big( \Rmg -i \sigma(f,g) \Rmg^2 \Rlf
\big) \Phi -  \Rmg \, \Phi \\
& =  -i \sigma(f,g) \, \Rmg^2 \Rlf \, \Phi \, .
\end{split}
\end{equation*}
Since the resolvent $\Rlf$ is bounded and has dense range, the space
$\Rlf \cD_{f,g,f}$ is
dense in $\cH$, and so the second part of the statement follows.
\end{Proof}

\noindent
\textbf{Remark:} It is an immediate consequence of the
second statement that the resolvents are analytic elements for the derivations.

\vspace*{2mm}
 The preceding lemma allows one to establish further properties of
the form $\sigma$.
Recall that $X$ is a real vector space, and by condition (I), the map
 $\delta:X\to{\rm Der} {\cal A}_0 $  is real linear.
Let $f,g,h \in X$,
$\kappa \in \RR \backslash \{0\} $ and $c \in \RR.$ Then
\begin{equation*}
\begin{split}
& \sigma(f+cg,h) \, R(\kappa, h)^2 \\
& = \delta_{f+cg}(R(\kappa, h)) = \delf(R(\kappa, h)) + c\,\delg(R(\kappa,
h)) \\
& = (\sigma(f,h) + c\,\sigma(g,h)) \, R(\kappa, h)^2 \, .
\end{split}
\end{equation*}
Hence $\sigma(f+cg,h) = \sigma(f,h) + c\,\sigma(g,h)$
and, by the antisymmetry of $\sigma$ established in
Lemma \ref{l2.5}, one also has
$\sigma(h,f+cg) = \sigma(h,f) +c\,\sigma(h,g)$.
Thus $\sigma$ is real linear in both entries.
Furthermore, if there is some $g \in X$ such that
$\sigma(f,g) = 0$  for all $f \in X$
one also has $\delf(\Rmg) = 0$, $f \in X$.
If $\delta$ acts ergodically on $\cA$ it follows that $g=0$
since otherwise $\Rmg$ is different from a multiple of
the identity. So the form $\sigma$ is non--degenerate in
this case. We summarize these results.

\begin{Proposition} \label{p2.7}
Let $(X,{\cal A})$ be
a flat \lcs, \ie $X$ is a real abelian Lie algebra and
the pair $(X,{\cal A})$ satisfies the conditions (I), (II)
and (III). Then there is an
antisymmetric bilinear form
$\sigma : X \times X \rightarrow \RR$ such that
for $\lambda, \mu \in \RR \backslash \{0\}, \ f,g \in X$
\begin{itemize}
\item[(i)]  \
$
[\Rlf, \Rmg] =  i \sigma(f,g) \, \Rlf \Rmg^2 \Rlf
$
\item[(ii)] \
 $\delta_f(\Rmg) = \sigma(f,g) \, \Rmg^2 \, . $
\end{itemize}
If $\delta$ acts ergodically, then $\sigma$ is
non--degenerate, \ie $(X,\sigma)$ is a symplectic space.
\end{Proposition}
These results show that the 
underlying C*--algebraic structure  determines a central
extension of the Lie algebra $X$ which is fixed by the form $\sigma$.
In a faithful irreducible representation
of ${\cal A}$ this extension of $X$ manifests itself
in the commutation relations of the generators of the
action $\delta$, given in Lemma \ref{l2.5}. The preceding proposition
expresses these relations in representation
independent C*--algebraic terms.

\section{Cohomology}
\setcounter{equation}{0}
\label{cohom}

In this section we continue our analysis of flat
\lcss{} and establish relations for pseudo--resolvents
which express additivity and homogeneity properties of the
underlying action $\delta$. Since the pseudo--resolvents are
not uniquely fixed by condition (II),
it is clear that we may have to
adjust them to establish such a result.
In the proofs we make use of standard  arguments
from cohomology theory, the main problem being
the control of domains of the generators in the chosen
Hilbert space representation of ${\cal A}$. For this, we  need the following
technical lemma.

\begin{Lemma} \label{l3.1}
Let $n \in \NN$, let
$f_1, \dots, f_n \in X$ and let $\lambda_1, \dots \lambda_n
\in \RR \backslash \{0\}$. The linear manifold
$\cD_{f_1, \dots, f_n} \doteq R(\lambda_n, f_n) \cdots R(\lambda_1, f_1) \, \cH$
is dense in $\cH$; it neither depends on the choice of
$\lambda_1, \dots, \lambda_n$ nor on the
particular order of $f_1, \dots, f_n$. Moreover,
$\cAo \, \cD_{f_1, \dots, f_n} \subset \cD_{f_1, \dots, f_n}$,
\ $\cD_{f_1, \dots, f_n} \subset \cD_{f_1, \dots, f_{n-1}}$
 and
$\cD_{f_1, \dots, f_n} \subset \bigcap_{\, k=1, \dots, n} \cD_{f_k}$
is a core for all operators $G_{f_k}$, $k = 1, \dots, n$.
\end{Lemma}
\begin{Proof}  The first part of the
statement follows by another application of the fact that
resolvents are bounded operators with dense range.
For the proof that
the parameters in $\RR \backslash \{0\}$ can be
arbitrarily chosen we use induction. The statement is
clear for $n=1$ so, in view of the induction hypothesis,
it suffices to show that $\lambda_{n+1}$ can be
replaced by any other parameter $\mu_{n+1}$ without changing the
respective domain.
To verify this one makes use of the resolvent equation
$$
R(\mu_{n+1}, f_{n+1}) = R(\lambda_{n+1}, f_{n+1})
\big( 1 + i(\lambda_{n+1} - \mu_{n+1}) \, R(\lambda_{n+1}, f_{n+1}) \big)
\, .
$$
Since $( 1 + i(\lambda_{n+1} - \mu_{n+1}) \, R(\mu_{n+1}, f_{n+1}))
\in \cAo$ by condition (III) one arrives,
by repeated application of Lemma \ref{l2.1},
at the inclusion
$$
R(\mu_{n+1}, f_{n+1}) R(\lambda_{n}, f_{n}) \cdots R(\lambda_1, f_1) \cH
\subset R(\lambda_{n+1}, f_{n+1}) R(\lambda_{n}, f_{n}) \cdots R(\lambda_1,
f_1) \cH \, .
$$
Interchanging $\lambda_{n+1}, \mu_{n+1}$ one obtains the opposite
inclusion, proving the independence of the domain on the choice of
parameters. For the proof that the order of the
chosen elements of $X$ does not matter either, it suffices to show
that one can permute $f_{n+1}$ and $f_n$ without changing the
domain. Now according to Lemma \ref{l2.6} and condition (III)
\begin{equation*}
\begin{split}
& R(\lambda_n, f_n) R(\lambda_{n+1},f_{n+1}) \\
& = R(\lambda_{n+1}, f_{n+1}) R(\lambda_n, f_n)
\big(1 - i \sigma(f_{n+1}, f_n) \,
R(\lambda_n, f_n) R(\lambda_{n+1}, f_{n+1}) \big) \\
& \doteq R(\lambda_{n+1}, f_{n+1}) R(\lambda_n, f_n) \, A_0 \, ,
\end{split}
\end{equation*}
where $A_0 \in \cAo$. By another application of Lemma \ref{l2.1}
it thus follows that
$$
R(\lambda_n, f_n) R(\lambda_{n+1},f_{n+1}) \cdots R(\lambda_1, f_1) \,
\cH \subset R(\lambda_{n+1},f_{n+1}) R(\lambda_n, f_n)  \cdots
R(\lambda_1, f_1) \,  \cH \, .
$$
According to the preceding step one can interchange $\lambda_n,\lambda_{n+1}$ in
this inclusion and interchanging also the role of $f_n, f_{n+1}$ one obtains the
opposite inclusion, proving equality. The proof of the independence
features of the domains $\cD_{f_1, \dots f_n}$ with regard to the
elements entering into their definition is therewith complete.

The stability of $\cD_{f_1, \dots f_n}$ under the action of $\cAo$
follows by still another application of Lemma~\ref{l2.1}. Hence, in
particular, $\cD_{f_1, \dots, f_n} \subset \cD_{f_1, \dots, f_{n-1}}$.
Finally, since $\cD_{f_1, \dots f_n} = R(\lambda, f_n) \, \cD_{f_1, \dots
  f_{n-1}}$ for arbitrary $\lambda \in \RR \backslash \{ 0 \}$, it is clear
that this domain is a core for $G_{f_n}$. But, as it is
invariant under permutations of the elements $f_1, \dots, f_n$,
it is a core for all generators $G_{f_k}$, $k = 1, \dots, n$.
\end{Proof}

Making use of this lemma we can establish the existence
of generators of the action~$\delta$ which are additive on $X$.

\begin{Lemma} \label{l3.1b}
There is a function $\gamma: X \rightarrow \RR$ such that
the ``improved'' generators
$$
\underline{G}_f \doteq \Gf - \gamma(f) \, 1 \, , \quad
f \in X \, ,
$$
are additive, \i.e.
\begin{equation} \label{3.2}
\big(\underline{G}_f + \underline{G}_g \big) \, \Phi
=   \,  \underline{G}_{f+g} \Phi \, , \qquad f,g \in X, \
\Phi \in   \cD_{f,g,f+g} \, .
\end{equation}
Moreover,  these generators have the same domain and
commutation properties as the given ones. The resulting resolvents
$$
\underline{R}(\lambda,f) \doteq (i\lambda \un + \underline{G}_f)^{-1}
= ((i\lambda - \gamma(f)) \un + \Gf)^{-1} \, ,
\quad f \in X, \ \lambda \in \RR \backslash \{0\} \, ,
$$
satisfy condition~(\ref{1.2}),
Lemma \ref{l2.6} and they are elements of $\cAo$.
\end{Lemma}
\begin{Proof}
Let
$f,g \in X$, $\Phi \in \cD_{f,g,f+g}$ and $A_0 \in  \cAo$. Then
$$
 [ \big(\Gf + \Gg - G_{f+g} \big), A_0] \, \Phi
 = -i \big(\delf(A_0) + \delg(A_0) - \delta_{f+g}(A_0)\big) \,   \Phi =
0 \,
$$
where we used Lemma \ref{l3.1} and Lemma \ref{l2.2}.
Thus by the generalized
Schur's lemmma there is a constant $\xi(f,g) \in \RR$, symmetric in
$f,g$,  such that
\begin{equation}
\label{3.1}
\big(\Gf + \Gg - G_{f+g} \big) \, \Phi
 = \xi(f,g) \,   \Phi \quad\hbox{for all}\quad\Phi\in\cD_{f,g,f+g}\,.
\end{equation}
Next, let $f,g,h \in X$ and pick any non--zero vector
$$
\Phi \in \cD_{f,g,h,f+g,g+h,f+g+h} \subset
\cD_{f,g,f+g} \cap \cD_{f+g,h,f+g+h} \cap \cD_{g,h,g+h} \cap \cD_{f,g+h,f+g+h} \, .
$$
Because of
the associativity of the addition of operators on a common
domain the preceding result entails
\begin{equation*}
\begin{split}
& \big(\Gf + \Gg + G_h \big) \, \Phi \\
& =  \big(G_{f+g} + G_h + \xi(f,g) 1 \big) \, \Phi =
G_{f+g+h} \, \Phi  +  (\xi(f,g) + \xi(f+g,h)) \, \Phi  \\
& = \big(G_{f} + G_{g+h} + \xi(g,h) 1 \big) \, \Phi =
G_{f+g+h} \, \Phi + (\xi(f,g+h) + \xi(g,h)) \, \Phi  \, .
\end{split}
\end{equation*}
Hence $\xi : X \times X \rightarrow \RR $ satisfies the cocycle equation
$$
\xi(f,g) + \xi(f+g,h) =  \xi(f,g+h) + \xi(g,h) \, ,
\quad f,g,h \in X \, .
$$
It is well known that for any abelian group $X$, all
real symmetric
solutions $\xi$ of this equation are coboundaries \cite{Ac}.
More concretely, for any such $\xi$
there is a function $\gamma: X \rightarrow \RR$ such that
$\xi(f,g) = \gamma(f) + \gamma(g) - \gamma(f+g)$, $f,g \in X$.
This is the $\gamma$ in the statement of the lemma, because
$
\underline{G}_f \doteq \Gf - \gamma(f) \, \un \, , \
f \in X \, ,
$
has the same domain and commutation properties as $\Gf$, and
$$
 \big(\underline{G}_f + \underline{G}_g \big)  \, \Phi
= G_{f+g}  \, \Phi
+  (\xi(f,g)-\gamma(f)-\gamma(g))  \, \Phi
=  \underline{G}_{f+g}\Phi
$$
for $f,g \in X$ and
$\Phi \in   \cD_{f,g,f+g}$. This establishes the claim (\ref{3.2}).
Since $\underline{R}(\lambda,f)=R(\lambda+i\gamma(f),f)$ is
contained in $\{R(z,f) :  z\in\C\backslash{i\R},\,f\in X\}$, the last claim
is also clear.
\end{Proof}

\vspace*{1mm}
The next result expresses the additivity property
(\ref{3.2}) in terms of the modified resolvents.
\begin{Lemma} \label{l3.2}
Let $f,g \in X$ and let $\lambda, \mu, (\lambda + \mu)
\in \RR \backslash \{ 0 \}$. Then
$$
\underline{R}(\lambda + \mu, f+g) \,
\big(\underline{R}(\lambda,f)  + \underline{R}(\mu,g)
+ i \sigma(f,g) \, \underline{R}(\lambda,f)^2
 \underline{R}(\mu,g)  \big) =
\underline{R}(\lambda,f)
\underline{R}(\mu,g) \, .
$$
\end{Lemma}
\begin{Proof}
Let $\Phi \in \cD_{f,g,f+g}$, then
\begin{equation*}
\begin{split}
& \underline{R}(\lambda + \mu, f+g) \,
\big(\underline{R}(\lambda, f) +  \underline{R}(\mu, g) \big)
\, \Phi \\
& = \underline{R}(\lambda + \mu, f+g) \, \underline{R}(\lambda,f)
\, \big(i(\lambda + \mu)1  +
\underline{G}_f  + \underline{G}_g \big)
\, \underline{R}(\mu, g) \, \Phi \\
& =  \big( \underline{R}(\lambda,f) \, \underline{R}(\mu, g)  +
 \underline{R}(\lambda + \mu, f+g) \,
[\underline{R}(\lambda,f), \underline{G}_{f+g}] \,
\underline{R}(\mu, g) \big) \, \Phi \\
& = \big( \underline{R}(\lambda,f) \, \underline{R}(\mu, g)  +
 \underline{R}(\lambda + \mu, f+g) \,
\big(i \sigma(g,f) \underline{R}(\lambda,f)^2 \big) \,
\underline{R}(\mu, g) \big) \, \Phi \, ,
\end{split}
\end{equation*}
where in the second equality relation (\ref{3.2}) was used
and in the third one Lemmata~\ref{l2.1} and \ref{l2.6}.
The statement then follows.
\end{Proof}

\vspace*{-3mm}
 By condition (I), the map
 $\delta:X\to{\rm Der} {\cal A}_0$  is linear,
 and this
raises the question whether there are underlying
generators which are not only additive but also homogenous on $X$.
An affirmative answer is given in the following lemma.

\begin{Lemma} \label{l3.2b}
There is a function $\vartheta: X \rightarrow \RR$ such that
the ``improved'' generators
$$
\overline{G}_f \, \Phi \doteq \big( \underline{G}_f -
\vartheta(f) 1 \big) \, \Phi \, , \quad
\Phi \in \cD_f \, .
$$
are real linear \i.e.
$$
\big(\overline{G}_f + c\,\overline{G}_g \big) \, \Phi \
  = \ \overline{G}_{f+cg} \, \Phi  \, , \qquad f,g \in X,\, c\in\RR,\,
\Phi \in   \cD_{f,g,f+cg} \, .
$$
and they have the same domain and commutation properties as $\Gf$.
The resulting resolvents
$$
\overline{R}(\lambda,f) \doteq (i\lambda 1 + \overline{G}_f)^{-1}
= \big((i\lambda - \gamma(f) - \vartheta(f))1 + \Gf \big)^{-1} \, .
$$
satisfy condition~(\ref{1.2}),
Lemma \ref{l2.6} and Lemma \ref{l3.2} and they are elements of
$\cAo$. 
\end{Lemma}
\begin{Proof}
Every vector space has a Hamel basis. Thus for $X$ there
is some index set $I$ and a subset
$\{h_\iota \in X : \iota \in I\}$, such that every element
$f \in X$ can be represented in a unique way as a finite
sum $f = \sum_\iota c_\iota^f h_\iota$ with coefficients
$c_\iota^f \in \RR$, $\iota \in I$. This basis will be kept fixed below.

We prove homogeneity in analogy to relation
(\ref{3.2}). Let
$f \in X$, $c\in\R$, $\Phi \in \cD_{f,cf}$ and $A_0 \in  \cAo$, then
by linearity of $\delta:X\to{\rm Der} {\cal A}_0$ we have
\[
 [ \big(c\,\Gf - G_{cf} \big), A_0] \, \Phi
 = \big(c\,\delf(A_0)  - \delta_{cf}(A_0)\big) \,   \Phi =
0 \, .
\]
Hence by the generalized Schur's lemma we get
that for given $\iota \in I$ and $c \in \RR$ there is some number
$\zeta_{\, \iota} (c) \in \RR$ such that
\begin{equation} \label{3.3}
\big( \underline{G}_{c h_\iota} - c \, \underline{G}_{h_\iota} \big)
\, \Phi
= \zeta_{\, \iota} (c) \,  \Phi \, , \quad \Phi \in
\cD_{h_\iota, \, c h_\iota} .
\end{equation}
Clearly $   \zeta_{\, \iota} (1) = 0$, and
 $  \zeta_{\, \iota} (0)  = 0$ since $\underline{G}_0=\zeta_{\, \iota}
 (0) \un$ on
 $\cD_{h_\iota}$, and $\underline{G}_f$ is additive in $f$.
Using this additivity, we also obtain for
 $c, c^\prime \in \RR$ and
$\Phi \in \cD_{h_\iota, \, ch_\iota, \, c^\prime \! h_\iota, \ (c + c^\prime) h_\iota}$
that
\begin{equation*}
\begin{split}
 \zeta_{\, \iota}(c + c^\prime) \, \Phi
& = \big( \underline{G}_{(c+c^\prime) h_\iota} -
(c + c^\prime) \, \underline{G}_{h_\iota} \big) \, \Phi \\
& = \big( \underline{G}_{c h_\iota} - c \, \underline{G}_{h_\iota}
+  \underline{G}_{c^\prime h_\iota} - c^\prime \,
\underline{G}_{h_\iota} \big) \Phi \\
& = \big(\zeta_{\, \iota}(c)  +  \zeta_{\, \iota}(c^\prime) \big) \Phi \, .
\end{split}
\end{equation*}
Hence $\zeta_{\, \iota} : \RR \rightarrow \RR$ is additive.
Now let $f \in X$ with corresponding
decomposition $f = \sum_\iota c_\iota^{\scriptscriptstyle f} h_\iota$. Since this
decomposition is unique and only a finite number of
coefficients $c_\iota^{\scriptscriptstyle f}$ are nonzero
we may define $ \vartheta(f) \doteq  \sum_\iota
\zeta_\iota(c_\iota^{\scriptscriptstyle f})$ which produces a map $\vartheta: X \rightarrow \RR$.
Let  $g = \sum_\iota c_\iota^{\scriptscriptstyle g} \, h_\iota\in X$, then
$f +g = \sum_\iota (c_\iota^{\scriptscriptstyle f} + c_\iota^{\scriptscriptstyle g}) \, h_\iota$
and so
\[
\vartheta(f+g)=\sum_\iota
\zeta_{\, \iota}(c_\iota^{\scriptscriptstyle f}
+c_\iota^{\scriptscriptstyle g}) = \sum_\iota
\big(\zeta_{\, \iota}(c_\iota^{\scriptscriptstyle f})
+\zeta_{\, \iota}(c_\iota^{\scriptscriptstyle g})\big)
=\vartheta(f)+\vartheta(g)
\]
hence $\vartheta$ is additive.
Since $ \underline{G}_f$ is also additive in $f$, it follows that the
 operators
$\overline{G}_f:\cD_f\to\cH$ given by
\begin{equation} \label{3.4}
\overline{G}_f \, \Phi \doteq \big( \underline{G}_f - \vartheta(f) \un
\big) \, \Phi \, , \quad \Phi \in \cD_f \, .
\end{equation}
are additive as well.
They have the same domain and commutation properties as $\Gf$.

For the proof that the generators
$\overline{G}(cf)$, $c \in \RR$, are homogenous
in $c$, consider first the case where $f=h_\iota$,
in which case $\vartheta(ch_\iota)=\zeta_{\, \iota}(c)$ and
$\vartheta(h_\iota)=\zeta_{\, \iota}(1)=0$.
Then
$$
\big( \overline{G}_{c h_\iota} - c \, \overline{G}_{h_\iota} \big)
\, \Phi
= \big( (\underline{G}_{c h_\iota} - \zeta_{\, \iota}(c) \, \un)
- c \, (\underline{G}_{h_\iota} - \zeta_{\, \iota}(1) \, \un ) \big) \, \Phi = 0
$$
making use of (\ref{3.3}) and
$\zeta_{\, \iota}(1) = 0$.
For the general case we make use of
the full power of Lemma~\ref{l3.1} for arbitrary $n \in \NN$.
Let $f = \sum_\iota c_\iota^{\scriptscriptstyle f}
h_\iota$.
Since only a finite number of the terms  $c_\iota^{\scriptscriptstyle f} h_\iota$
is different from zero there is some dense domain
$\cD \subset \cD_{f, cf}$
which is stable under the action of $\cA_0$ and
it lies in the domains of all generators $G_k$ with
$k\in\{ c_\iota^{\scriptscriptstyle f} h_\iota, c
c_\iota^{\scriptscriptstyle f} h_\iota : \iota \in I \} $
as well as $k$ being a sum of these.
Hence one obtains for $\Phi \in \cD$,
\begin{equation*}
\begin{split}
\big( \overline{G}_{cf} - c \, \overline{G}_{f} \big) \, \Phi
& = \big(\overline{G}_{\sum_\iota c c_\iota^{\scriptscriptstyle f} h_\iota}
-  c \, \overline{G}_{\sum_\iota c_\iota^{\scriptscriptstyle f} h_\iota} \big) \, \Phi
\\
& = \sum_\iota \big(\overline{G}_{c c_\iota^{\scriptscriptstyle f} h_\iota}
- c\, \overline{G}_{c_\iota^{\scriptscriptstyle f} h_\iota} \big) \, \Phi
= \sum_\iota \big(c c_\iota^{\scriptscriptstyle f} \overline{G}_{h_\iota}
- c \,c_\iota^{\scriptscriptstyle f} \overline{G}_{h_\iota} \big) \, \Phi = 0 \, ,
\end{split}
\end{equation*}
where in the second equality the additivity of $\overline{G}$ was used
and in the third equality homogeneity w.r.t. $h_\iota$. Since $\cD$ is a core
for the underlying generators, it follows that
$\overline{G}_{c f}  =  c \, \overline{G}_{f}$ on
$\cD_{f, \, cf}$.

As before we define the improved resolvents
$$
\overline{R}(\lambda,f) \doteq (i\lambda \un + \overline{G}_f)^{-1}
= \big((i\lambda - \gamma(f) - \vartheta(f)) \un + \Gf \big)^{-1} 
$$
which still satify condition~(\ref{1.2}) and 
Lemma \ref{l2.6}.  
Since the generators
$\overline{G}$ are additive, the corresponding resolvents also satisfy the
relation given in Lemma \ref{l3.2}.
The last claim is clear in view of condition (III).
\end{Proof}

\vspace*{1mm}
The homogeneity of the
generators manifests itself in further
algebraic properties.

\begin{Lemma} Let $f \in X$ and let $\lambda, c \in \RR \backslash \{ 0 \} $.
Then
$$
c \, \overline{R}(c \lambda, c f) = \overline{R}(\lambda, f) \, .
$$
\end{Lemma}
\begin{Proof}
Pick any vector $\Phi \in \cD_{f,cf}$, then it follows from
Lemma \ref{l3.1} that
\begin{equation*}
\begin{split}
& \big( c \, \overline{R}(c \lambda, c f) - \overline{R}(\lambda, f)
\big) \, \Phi \\
& =
\overline{R}(c \lambda, c f) \,
\big(c(i \lambda \un + \overline{G}(f) ) -
(i c \lambda \un + \overline{G}(c f) ) \big)
\overline{R}(\lambda, f) \,  \Phi = 0 \, ,
\end{split}
\end{equation*}
where the second equality follows from the homogeneity
of the generators. \end{Proof}

We summarize our findings. Let $\cR \subset \cA$ be
the C*--algebra generated by the range of the underlying
resolvents $R$. In the preceding discussion we have
shown that one can proceed from these resolvents by
analytic continuation to improved resolvents
$\overline{R} \in \cR$, in which the
vector space structure of $X$ manifests itself by
additional relations. We have worked in a concrete
representation of $\cA$. But since this representation
was faithful the above  results can be reformulated in the
abstract setting.

\begin{Theorem}
Let $(X,{\cal A})$ be a flat \lcs, \ie $X$ is a
real abelian Lie algebra and the pair
$(X,{\cal A})$ satisfies the conditions (I), (II) and (III).
Let $\cR \subset \cA$ be the C*--algebra
generated by the corresponding pseudo--resolvents.
There are a skew--symmetric bilinear form
$\sigma : X \times X \rightarrow \RR$ and 
pseudo--resolvents $\{\overline{R}(\lambda, f)
 : \lambda \in \RR \backslash \{ 0 \}, \ f \in X
 \}\subset\cR\bigcap\cA_0$  such that
 \[
 \overline{R}(\lambda, f) \, \delf(A_0) \, \overline{R}(\lambda, f)
  = i \, [A_0 , \overline{R}(\lambda, f)] \, , \quad A_0 \in \cAo \, ,
 \]
and, for $f,g \in X$,
$\lambda, \mu \in \RR \backslash \{0\}$ and $\lambda + \mu \neq 0$ in
item (ii),
\begin{itemize}
\item[(i)]  \
$
[ \overline{R}(\lambda, f),  \overline{R}(\mu, g)] =  i \sigma(f,g) \,
\overline{R}(\lambda, f) \overline{R}(\mu, g)^2 \, \overline{R}(\lambda, f)
$
\item[(ii)] \  $ \overline{R}(\lambda + \mu, f+g) \,
\big(\overline{R}(\lambda,f)  + \overline{R}(\mu,g)
+ i \sigma(f,g) \, \overline{R}(\lambda,f)^2\,
 \overline{R}(\mu,g)  \big) = \overline{R}(\lambda,f) \overline{R}(\mu,g)
$
\item[(iii)] \
$c \, \overline{R}(c \lambda, c f) = \overline{R}(\lambda, f)$
for $ c \in \RR \backslash \{ 0 \} $
\item[(iv)] \
 $\delta_f(\overline{R}(\mu,g)) = \sigma(f,g) \, \overline{R}(\mu,g)^2
 \, , \quad \mu  \in \RR \backslash \{0\} \, $.
\end{itemize}
If $\delta$
acts ergodically on $\cA$, then $\sigma$ is non--degenerate.
\end{Theorem}

As the range of the analytic continuations of the resolvents is contained in
$\cR$, this algebra is generated by the improved resolvents as well.
Moreover, if $\delta$ 
acts ergodically,  the relations obtained above show that
the algebra $\cR$ is just the resolvent algebra $\cR(X,\sigma)$,
defined in \cite{BuGr1}. It is noteworthy that $\cR(X,\sigma)$  is
primitive since its Fock representation is faithful.
Finally, we show that in general 
the algebra $\cR$ is independent from the choice
of pseudo--resolvents satisfying relation (1.2).

\begin{Proposition}
Let $(X,{\cal A})$ be a flat \lcs{} and let
$\{\Rlf \in \cA_0: \lambda \in \RR \backslash \{ 0 \}, f \in X \}$ and
$\{\Rplf \in \cA_0: \lambda \in \RR \backslash \{ 0 \}, f \in X \}$
be two families of pseudo--resolvents satisfying relation (\ref{1.2}).
Then the respective C*--algebras generated by these families
coincide. \end{Proposition}
\begin{Proof}
As  before, assume without loss of generality that we
have concretely $\cA\subseteq \cBH$ for some
Hilbert space $\cH$ and $\cA^- = \cBH$, where the bar denotes
weak closure. Thus there are selfadjoint generators $\Gf$ (resp. $\Gpf$)
which are densely defined on the domain $\cD_f \doteq
\Rlf \cH$ (resp. $\cD^{\, \prime}_f \doteq
\Rplf \cH$) and satisfy $\Rlf = (i\lambda \un + \Gf)^{-1}$
(resp. $\Rplf = (i\lambda \un + \Gpf)^{-1}$).

Let $\cD \doteq \Rlf \Rplf \, \cH$
for fixed $\lambda \in \RR \backslash \{0\} $, $f \in X $.
Since resolvents are bounded and their range is dense, it is clear that
$\cD$ is dense in $\cH$. By Lemma~\ref{l2.1}, we also have that
for any $A_0 \in \cAo$ there is a  $B_0 \in \cAo$
such that $A_0 \Rlf = \Rlf B_0$ and also the analogous statement
for $\Rplf$. Thus we obtain stability of $\cD$ under the action
of $\cAo$. By definition
$\cD \subset \cD_f$ is a core for $\Gf$.
If we let $A_0 = \Rlf$ then $\Rlf \Rplf = \Rplf B_0$ for 
some $B_0\in\cAo$, hence
 $\cD \subset \Rplf \, \cH = \cD^{\, \prime}_f$.
Moreover, as $\Rplf \in \cAo$ it is also clear that
$\Rplf \, \cD \subset \cD$, and hence $\cD$ is a core
for $\Gpf$ as well.
Now  observe that for $\Phi \in \cD$
and $A_0 \in \cAo$ we have via Lemma~\ref{l2.2} that
$$ [(\Gpf - \Gf), A_0 ] \, \Phi =
-i (\delta_f(A_0) - \delta_f(A_0)) \, \Phi = 0 \, . $$
Thus by the generalized Schur's lemma there is some $c_f \in \RR$ such that
$(\Gpf - \Gf - c_f \un) \upharpoonright \cD = 0$. Since $\cD$
is a core for $\Gf$ and $\Gpf$ it follows that
$\Gpf = \Gf +  c_f \un$, proving that
$$\Rplf = \big((i \lambda + c_f) \un + \Gf)^{-1}
\in \{R(z,f) :  z\in\C\backslash{i\R} \}
\subset \cR \,
. $$
Thus the C*--algebra generated by 
$\{\Rplf \in \cA_0 : \lambda \in \RR \backslash \{ 0 \}, f \in X \}  $
is contained in the C*--algebra generated by 
$\{\Rlf \in \cA_0 : \lambda \in \RR \backslash \{ 0 \}, f \in X \}$
and by symmetry of the argument we also have the reverse inclusion, hence equality.
\end{Proof}

\section{Concluding remarks}
\setcounter{equation}{0}

We have established above a C*--algebraic framework for
the systematic study of the
representation theory of Lie--algebras of derivations tailored to the needs of quantum
physics. For the simple case of flat \lcss{}
$(X,{\cal A})$ arising from actions of abelian Lie
algebras $X$ on primitive C*--algebras ${\cal A}$, we
were able to completely determine the algebraic structure of the
generators. It turned out that this structure provides
in general a central extension of $X$ whose specific form
is fixed by some skew--symmetric bilinear form
$\sigma : X \times X \rightarrow \RR$
encoded in the underlying algebraic data. Remarkably, the C*--algebra generated
by the resolvents of the generators coincides with the
resolvent algebra $\cR(X,\sigma)$, invented in \cite{BuGr1}
as a convenient framework for the description of quantum systems.

In view of these results it seems worthwhile to extend this
study of representations of Lie algebras of derivations
to the non--abelian case. It has to be noted that the technical
condition (III) would no longer be meaningful in this
general context, \ie the
pseudo--resolvents do not need to belong to the
domain of the action $\delta$. As a matter of fact, as these
pseudo--resolvents are assumed
to be elements of ${\cal A}$, cf.\ condition (II),  one may even have
to relax the assumption that the domain  ${\cal A}_0$
of $\delta$ is norm dense in ${\cal A}$. It would still be
meaningful to require that this domain is weakly dense in
all faithful representations of ${\cal A}$. In fact, the
present results can be established under this weaker
assumption. A solution of these mathematical
problems would be rewarding since it would
shed new light on the appearance of central
extensions of symmetry groups in the context of quantum physics.

\section{Acknowledgements}

DB and HG are grateful for the support of the Courant Research 
Center ``Higher Order Structures'' of the University of G\"ottingen


\begin{thebibliography}{99}
\bibitem{Ac} J.\ Acz\'el, The general solution of two functional
equations by reduction to functions additive in two variables and
with the aid of a Hamel basis,
{\sl Glasnik Mat.--Fiz. Astronom., \bf 20}, 65--73 (1965).


\bibitem{BuGr1} D.\ Buchholz and H.\ Grundling,
The Resolvent Algebra: A New Approach to Canonical Quantum Systems.
J.\ Funct.\ Analysis {\bf 254}, 2725--2779 (2008)

\bibitem{BuGr2} D.\ Buchholz and H.\ Grundling, 	
Algebraic Supersymmetry: A Case study.  Commun.\
Math.\ Phys.\ {\bf 272}, 699--750  (2007)

\bibitem{Dix}
Dixmier, J.: C*-algebras,
North Holland Publishing Company, Amsterdam - New York - Oxford 1977


\bibitem{Ka59}
Kato, T.: Remarks on Pseudo-resolvents and Infinitesimal
Generators of Semi-groups.
Proc. Japan Acad. {\bf 35} (1959), 467--468.

\bibitem{Sa} S.\ Sakai, \textit{C*--algebras and W*--algebras},
Springer (1971)


\bibitem{Yo} K.\ Yosida, \textit{Functional Analysis}, 
Springer (1980)

\end{thebibliography}
\end{document}